# Nanoscale imaging of antiferromagnetic domains in epitaxial films of Cr$_2$O$_3$ via scanning diamond magnetic probe microscopy

*Adam Erickson,[‡] Syed Qamar Abbas Shah,[§] Ather Mahmood,[§] Ilja Fescenko,[∥] Rupak Timalsina,[‡] Christian Binek,[§,*] and Abdelghani Laraoui[‡,§,&]*

[‡]Department of Mechanical & Materials Engineering, University of Nebraska-Lincoln, 900 N 16th St, W342 NH, Lincoln, Nebraska 68588, USA

[§]Department of Physics and Astronomy and the Nebraska Center for Materials and Nanoscience, University of Nebraska-Lincoln, 855 N 16th St, Lincoln, Nebraska 68588, USA

[∥]Laser Center, University of Latvia, Jelgavas St 3, Riga, LV-1004, Latvia

**Abstract**
We report direct imaging of boundary magnetization associated with antiferromagnetic domains in magnetoelectric epitaxial Cr$_2$O$_3$ thin films using diamond nitrogen vacancy microscopy. We found a correlation between magnetic domain size and structural grain size which we associate with the domain formation process. We performed field cooling, *i.e.*, cooling from above to below the Néel temperature in the presence of magnetic field, which resulted in the selection of one of the two otherwise degenerate 180$^0$ domains. Lifting of such a degeneracy is achievable with a magnetic field alone due to the Zeeman energy of a weak parasitic magnetic moment in Cr$_2$O$_3$ films that originates from defects and the imbalance of the boundary magnetization of opposing interfaces. This boundary magnetization couples to the antiferromagnetic order parameter enabling selection of its orientation. Nanostructuring the Cr$_2$O$_3$ film with mesa structures revealed reversible edge magnetic states with the direction of magnetic field during field cooling.

**1. Introduction**
Under zero applied magnetic field and below Néel ordering temperature ($T_N$) the antiferromagnetic (AFM) interactions in AFM materials lead to a collinear or noncollinear spin orientation in the ground state that can be represented by one single vector of unit length, called the Néel vector.[1] Because of their zero net magnetization, AFMs were originally considered less useful for spintronic devices. This presumption has been negated by recent discoveries in the electrical control and detection of the orientation of the Néel vector in high $T_N$ (> 300 K) AFMs materials such as CuMnAs,[2–4] Mn$_2$Au,[5,6] and Mn$_3$Sn.[7,8] In addition, these materials are robust against external magnetic fields and display no magnetic stray fields. The potential for ultrafast dynamics in the THz range makes antiferromagnets well suited for miniaturized ultrafast spintronic devices.[9,10]

Magnetoelectric (ME) AFMs have an equilibrium surface or boundary magnetization that couples to Néel vector and can be controlled by electric field, offering additional means for controlling magnetic order of AFM materials for spintronics devices.[11] The AFM ME sesquioxide Cr$_2$O$_3$ (chromia) is an archetypical oxide that permits voltage-control of the Néel vector in the presence of an applied magnetic field.[11,12] Thin films of Cr$_2$O$_3$ have been used to realize voltage-control of the peculiar boundary magnetization of single domain ME AFMs, detected through an exchange bias produced by Cr$_2$O$_3$ on an adjacent ferromagnet CoPd.[12] A purely antiferromagnetic magnetoelectric random access memory has been realized in Pt(20 nm)/α-Cr$_2$O$_3$(200 nm)/Pt(2.5 nm) structures with 50-fold reduction of the writing threshold compared with ferromagnet-based materials,[13] making chromia a potential material to use in AFM spintronics.[14] Later magnetic force microscopy (MFM) and photoemission electron microscopy (PEEM) combined with x-ray magnetic circular dichroism (XMCD) were used to spatially map the electrically controlled surface magnetization domain structures of pristine $Cr_2O_3$ films with domain size of few micrometers. However, the MFM and XMCD-PEEM contrast is very small due to the weak stray magnetic field generated from uncompensated spins at the surface of $Cr_2O_3$ film,[15] and the used techniques tend to suffer from low spatial resolution (e.g., > 50 nm in PEEM).[16,17]

Magnetic microscopy based on nitrogen vacancy (NV) centers in diamond has become a versatile tool to study magnetic phenomena at the nanoscale.[18–22] NV scanning probe magnetometry (NV-SPM) magnetic imaging of 180-degree domains in granular thin films of $Cr_2O_3$ revealed an average magnetic domain size of 230 nm with narrow domain walls[23] contrary to theoretical expectation based on the weak anisotropy of $Cr_2O_3$.[24] NV-SPM in combination with second-harmonic-generation microscopy measurements on bulk $Cr_2O_3$ crystals showed that most $180^0$ domain walls are Bloch-like and can coexist with Néel walls in crystals with a high in-plane anisotropy.[25] The domain wall (DW) width has important implications for antiferromagnetic spintronic device applications. Large DWs limit device scaling and investigations in epitaxial $Cr_2O_3$ films are missing. Here we report NV-SPM imaging of surface AFM domains with DWs and edge domains of nanostructured epitaxial $Cr_2O_3$ (0001) thin films under different experimental (magnetic field, temperature) conditions in thin films grown via pulsed laser deposition. We found distinct differences in the magnetic domain formation compared with reports in the literature on samples grown by sputtering methodology.

## 2. Experimental conditions

The samples used in this study consist of $Cr_2O_3$ films grown on (0001) sapphire substrates. The sapphire substrates were cleaned using a modified Radio Corporation of America protocol.[26] 200 nm thick $Cr_2O_3$ films were deposited using pulsed laser deposition (PLD).[27] The substrates were heated to 1073 K during the deposition. A KrF excimer laser with pulse energies of 200 mJ, a spot size of about 6 mm$^2$, and a pulse width of 20 ns (at a repetition rate of 10 Hz) was used to ablate the $Cr_2O_3$ target. The target-to-substrate distance was kept at about 9 cm and the substrate rotation rate was at 4 rpm. X-ray diffraction (XRD) measurements (**Fig. 1a**) revealed the (0001) orientation of $Cr_2O_3$ film. The Atomic force microscopy topography map (**Fig. 1b**) shows the root mean square (RMS) surface roughness value of 0.155 nm (0.13 nm in the dashed square area).

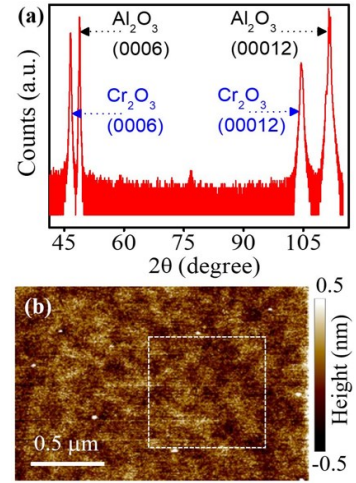

**Figure 1**. **(a)** XRD pattern showing the $Cr_2O_3$ (0006) and $Al_2O_3$ (00012) peaks. **(b)** Atomic force microscopy topography image of the $Cr_2O_3$ (0001) film surface.

The NV-SPM microscope (**Fig. 2a**) used in this study is a state-of-the-art home built combined platform hosting tuning-fork-based atomic force microscope and optical confocal microscope.[28] We used a long working distance (4 mm), high NA (= 0.8) Nikon objective and a Qnami probe tip. The tuning fork is attached to Attocube XYZ step motors with positioning resolution below 1 nm. An additional piezo-based xyz scanner, purchased from Npoints, provides a scan range of 100 μm×100 μm x 20 μm. The special self-sensing and self-actuating SPM sensor is based on a diamond cantilever attached to a quartz rod (**Fig. 2d**). **Fig. 2c** shows the fluorescence image of NV center implanted 9 ± 3 nm below the surface of a (100) oriented diamond probe (**Fig. 2b**). We used SRIM calculations to determine the NV depth for $^{15}$N implantation of diamond substrate with an energy of 5 keV.[29] The negatively charged NV center, composed of a substitutional nitrogen near to a vacancy site, is an electronic spin 1 with a spin-triplet ($|m_s = 0>$, $|m_s = \pm 1>$) in the ground state. 532-nm laser illumination induces spin-dependent fluorescence (650- 800 nm)[18] allowing optical detected magnetic resonance (ODMR) of its spin state, **Fig. 2b**. The applied magnetic field, provided by a permanent magnet (up to 30 mT), splits $|m_s = \pm 1>$ state via Zeeman effect and leads to two ($|m_s = 0>$ to $|m_s = -1>$ and $|m_s = 0>$ to $|m_s = -1>$) ODMR peaks whose frequencies depend on the projection of the field along the NV symmetry axis. The setup is integrated with microwave (MW) equipment to monitor NV spin transitions and with a single phonon counter module coupled with a single mode fiber.[30–32]

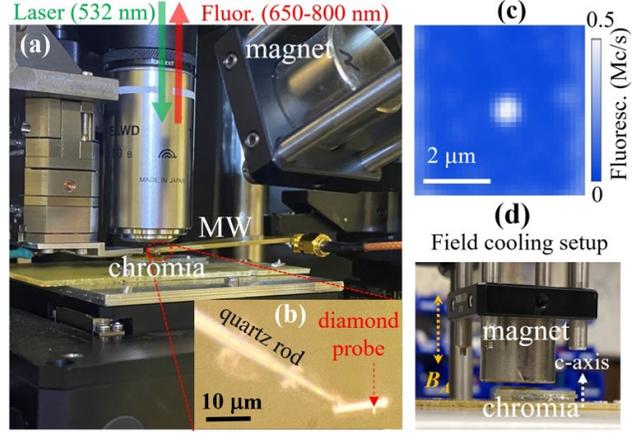

**Figure 2.** (a) A picture of NV-SPM setup showing 532-nm laser excitation and fluorescence (650-800 nm) detection through a high NA objective. (b) Optical image of the diamond probe attached to a quartz rod. (c) Fluorescence image of NV located ~12 nm from the diamond probe surface in (b). (d) Field cooling setup with a permanent magnet (up to ± 1 T) mounted on a hot plate (up to 500 K).

## 3. Results and discussion

NV magnetometry probes the stray magnetic field resulting from local surface spins[19–21]. In noncollinear AFMs, the measured stray field emerges from an overall ferrimagnetic moment due to the canting of the spins.[8,33,34] However in collinear AFMs such as $Cr_2O_3$, NV magnetometry senses the stray field $B_{NV}$ generated from a single layer of uncompensated surface spins, **Fig. 3a**.[23,25,35] We performed ODMR measurements on the diamond probe in **Fig.2c** by sweeping the MW frequency across the NV $|m_s = 0>$ to $|m_s = -1>$ transition at an applied field $B_A$. The Hamiltonian of the system in the lab coordinates (x, y, z) in **Fig. 3a** is:[18,36]

$$H = DS_z^2 - \gamma_{NV}\big(S_x(B_{Ax} + B_x) + S_y(B_{Ay} + B_y) + S_z(B_{Az} + B_z)\big),$$

Where $D$ is the zero-field splitting = 2.87 GHz, $\gamma_{NV}$ = 28 GHz/T is the gyromagnetic ratio of the electron spin. The second term of the Hamiltonian is the Zeeman splitting term. By increasing /decreasing the amplitude of $\boldsymbol{B_A}$ the resonance of the $|m_s = 0>$ to $|m_s = -1>$ transition peak shift to higher/lower frequencies respectively. By monitoring the NV fluorescence increase/decrease, we can measure the magnetic field generated by the magnet $B_A$ and the additional local stray field $B_{NV}$ generated from scanning across the spin textures of $Cr_2O_3$ film. This is the basis of DC magnetic sensing scheme of NV magnetometry.[18–21] The optimized DC measurable magnetic field in the ideal photon-shot-noise limit is given by:[22,36]

$$B_{min} \cong 4\,\Gamma\,(3\sqrt{3}\,\gamma_{NV}\,C)^{-1}\,(I_0 t)^{-1/2},$$

where $\Gamma$ is the full-width-at-half-maximum linewidth of the ODMR peak, $C$ is the ODMR peak contrast, $I_0$ is the NV fluorescence rate, and $t$ is the measurements time.[37] By using the parameters of the NV measurements in **Fig. 2c** and **Fig. 3b** ($I_0$ = 500k counts/s, $\Gamma$ = 7.78 MHz, C = 0.06,) we found $B_{min}$ = 5 µT for $t$ = 1s and a confocal detection voxel of 350 x 350 nm$^2$, which is more than sufficient to resolve magnetic stray fields $B_{NV}$ from AFM domains in $Cr_2O_3$ (**Fig. 3c**).

For our measurement we used $B$ dual-Iso imaging method[38] to image AFM domains in 200-nm thick $Cr_2O_3/Al_2O_3$ substrate. The NV fluorescence is recorded alternatingly at two MW frequencies centered around the ODMR frequency of transition $|m_s = 0 >$ to $|m_s = -1 >$, chosen at the maxima of the derivative of the measured ODMR peak (crossed dots in the insert of **Fig. 3b**). By measuring the NV fluorescence (*Fl*) at each point we can measure directly the signal $S = (Fl(f_1)-Fl(f_2))/(Fl(f_1)+Fl(f_2))$, which gives positive or negative magnetic stray fields based on the spins orientation of the magnetic domain.[28,39] This method is suitable for imaging magnetic samples with high (> 10 µT) stray fields and allows faster scanning times in comparison to recording the full ODMR spectrum.[40] The obtained $B_{NV}$ image at an applied filed $B_A$ = 1.07 mT shows two oppositely magnetized domains with areas of positive $B_{NV}$ (red) and negative (blue) $B_{NV}$, separated by sharp domain walls zeros in $B_{NV}$ (white color), **Fig. 3c**.

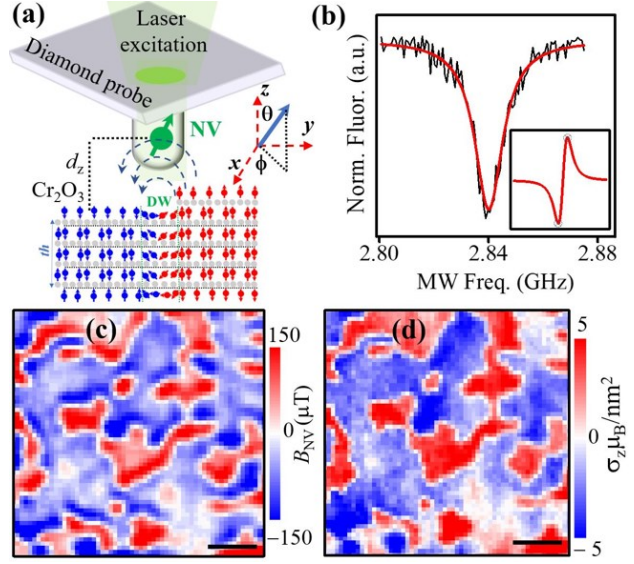

**Figure 3. (a)** A sketch of NV-SPM with an NV located a distance $d_z$ from $Cr_2O_3$ film surface with opposite magnetic domains separated by a domain wall. Strong magnetic stray fields (dashed field lines) are generated at the DW indicating regions of opposite order parameter defined by the orientation (red up or blue down) of the $Cr^{3+}$ atom. **(b)** ODMR peak on NV at $B_A$ = 1.07 mT fitted with Lorentzian function. Insert: zoomed derivative spectrum of an ODMR peak for $|m_s = 0 >$ to $|m_s = -1 >$ transition to select the two MW points needed for $B$ dual-Iso imaging. **(c)** $B_{NV}$ map of 5-µm $Cr_2O_3$ region (scale bar is 1 µm). **(d)** The extracted moment density profile of image in **(c)** which reveals a domain pattern of spin-up and spin-down domains (scale bar is 1 µm).

In most of surface magnetometry techniques, the inverse problem is used to calculate the magnetization distribution.[21] In here, we used a similar model to references [23] and [41] of $B_{NV}$ by describing the magnetization $m$ of the AFM film as two monolayers of out-of-plane spins (Fig. 2a) with opposite directions, separated by a distance $th$ and have a moment density $\sigma_z(x, y)$:[23] $m(x, y, z) = \sigma_z(x, y)[\delta(z) - \delta(z + th)]z$, where $th$ is the thickness of the chromia film = 200 nm, $\delta$ is the Dirac delta function and $z$ is the unitary direction. The calculated $B_{NV}$ is then obtained by field propagation in the Fourier space:[23,25,41] $B_{NV}(\vec{k}) = m(\vec{k}) T_{NV}(d_z, \theta_{NV}, \phi_{NV}, \vec{k})$, where $T_{NV}$ is a propagator that depends on the NV orientation ($\theta_{NV}, \phi_{NV}$) and the NV-to-sample distance $d_z$ (**Fig. 3a**). The magnetic moment density profile is determined from the measured $B_{NV}$ map using:[23,41] $\sigma(k) = T_{NV}^{-1}(d_z, \theta_{NV}, \varphi_{NV})W(k)B_{NV}(k)$, where $W(k)$ is the filter function given by a Hanning window in the Fourier space.[23] By using the parameters of our NV geometry ($d_z$ = 50 nm, $\theta_{NV}$ = 54°, and $\phi_{NV}$ = 92°), we reverse propagated the measured $B_{NV}(x, y)$ map in **Fig. 3c** to calculate $\sigma_z(x, y)$ map displayed in **Fig. 3d**. The resulting calculated magnetization image shows the presence of homogeneously magnetized domains, with well-defined domain walls with surface moment density of ~ 5 $\mu_B/nm^2$. These regions correspond to antiferromagnetically ordered states with opposing orientation of the Néel vector, as sketched in **Fig. 3a**.

The size of the measured domains varies between 200 nm to 2 µm and it is bigger than the average AFM domain size measured on granulated films (~ 230 nm).[23] We attribute this finding to the fact that our samples have been grown by PLD. Previously,[42] sister samples grown under identical conditions via PLD have been characterized via high angle annular dark-field (HAADF) scanning transmission electron microscopy (STEM). The STEM data revealed epitaxial growth of $Cr_2O_3$ (0001) on *c*-plane sapphire. Although PLD grown samples exhibit a new type of planar crystallographic defects on length scales of the order of just a few nanometers, which are comprised of a 60° rotation about the *c*-axis combined with a 1/3 $[10\bar{1}0]$ lattice shift,[42] these defects leave the magnetic exchange interaction between $Cr^{3+}$ ions and the resulting antiferromagnetic order virtually unaffected. Grains and their grain boundaries in sputtered samples behave differently. Here exchange bonds are broken turning grain boundaries into magnetic defects. The abundance of grains in sputtered samples increases the number of domain nucleation sites. With an increased formation of nucleation sites on cooling to below the Néel temperature, the likelihood that two growing antiferromagnetically ordered regions with opposing orientation of the Néel vector meet and form a domain wall increases. With an increasing number of magnetic domains their average size decreases accordingly. This mechanism explains the reduced domain size in sputtered samples with grains on the order of 50 nm,[23] while also providing a statistical interpretation of the fact that magnetic domains can and typically do grow beyond the size of the grains.

The antiferromagnetic domain pattern in $Cr_2O_3$ film is strongly affected by the applied magnetic field and temperature conditions, as shown in **Fig. 4**. To demonstrate this we field-cooled the chromia film (heating to 320 K (>$T_N$ = 307 K[12]) and cooling to ambient condition (*T* = 295 K) under applied parallel (positive)/anti-parallel (negative) magnetic field $B_A$ applied along the c-axis of the chromia film, **Fig. 2d**. Prior to field cooling we imaged the film and recorded the $B_{NV}$ map in **Fig. 4a** with a similar domain pattern to **Fig. 3c**. The result of the field cooling at $B_A$ = -0.4 T is an ordered spin state with no apparent AFM domains (**Fig. 4b**). The experimental confirmation of selection of a single domain state confirms that in thin films where parasitic magnetization is present, magnetic field cooling in the absence of an applied electric field can select an antiferromagnetic single domain state with uniform boundary magnetization.[43,44]

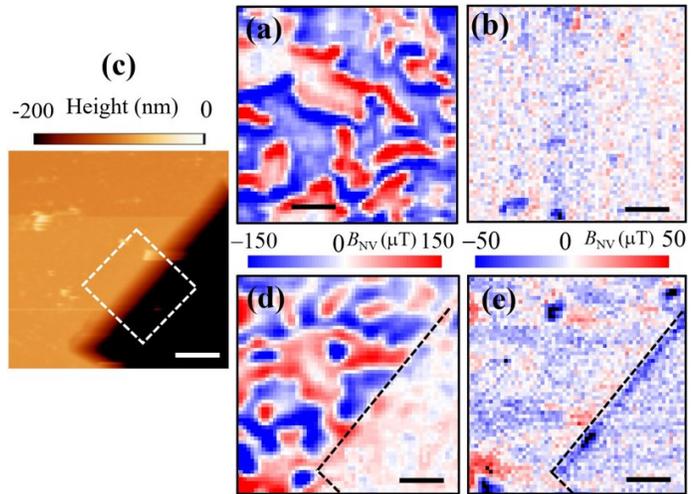

**Figure 4. Magnetic field cooling of AFM magnetic domains in chromia.** (a) $B_{NV}$ map of a given region in chromia film prior to filed cooling. (b) $B_{NV}$ map of region (a) after heating above 320 K then cooling to 295 K under a magnetic field of -0.4 T. (c) Atomic force microscope topography map of a zoomed FIB etched nanostructure in $Cr_2O_3$ film. $B_{NV}$ map of the region shown in (c) prior (d) and after (e) field cooling under a magnetic field of -0.4 T. Scale bar is 1 µm in (a), (b), (d), and (e).

Previous experiments were based on integral magnetometry and the presence of uniform boundary magnetization was an interpretation rather than a measured observation. In contrast to polydomain states where higher stray fields are observed in **Fig. 4b** at the DW edges the $B_{NV}$ contrast is low due to the weak stray-field generated from the out-of-plane component of the surface magnetization. This can also be seen at the center of the AFM domains in **Fig. 3c** and **Fig. 4a**, far away from the DW edges. To obtain a measurable magnetic stray field from uniformly ordered magnetic domains,[35] we made micron-scale mesas by etching features in $Cr_2O_3$ film grown on $Al_2O_3$ by using focused ion beam (FIB, provided by FEI Helios NanoLab 660). We used a voltage of 30 kV and a current of 7.7 pA to etch 200 nm down the chromia film. We performed scanning electron microscopy (SEM) imaging, not shown here, to monitor the etching and image the etched film regions. **Fig. 4c** shows the topography atomic force microscope image of a zoomed 5 μm x 10 μm FIB structure with a depth of ~ 200 nm. The dark area corresponds to the $Al_2O_3$ substrate. We first imaged the pristine region in **Fig. 4c** and found the presence of the AFM domains in the unetched region and no apparent contrast is present in the etched region (**Fig. 4d**) as expected from $Al_2O_3$. We then used a similar field cooling as in **Fig. 2d** and observed only a magnetic contrast (blue: negative stray field) at the edge of the etched regions, **Fig. 4e**. Some of the AFM domains in the unetched regions of the film are not switched completely as in **Fig. 4b** that we explain by the FIB induced defects resulting in pinned magnetic domains.[45]

The switching of AFM order is clearly seen at the edge of the etched region where the FIB edge defects create a residual edge magnetic domain with a stronger out of plane stray field in comparison to the unetched region (**Fig. 4b**). The reason of the AFM order switching in the presence of only applied magnetic field is that in thin $Cr_2O_3$ films grown by any method (*e.g.*, PLD in our case) there is a parasitic magnetic moment that originates from defects in the bulk and the difference in the boundary magnetization between the surface next to vacuum and the surface with the substrate (sketch in **Fig. 3a**). This magnetic moment is tied to the orientation of the antiferromagnetic order parameter.[15] Because the magnetic moment orients with the direction of magnetic field on cooling (**Fig. 5**), the Néel vector orients accordingly. In bulk $Cr_2O_3$ crystals the parasitic moment is very weak, and the Zeeman energy induced by the applied magnetic field is not strong enough to switch the orientation of Néel vector. One need to apply an electric field in addition to the applied magnetic field to activate the magnetoelectric effect and thus creating an energetic favor of one AFM single domain over the other.[12,13,15,35]

To determine the spatial profile of the nanostructured edge AFM state and whether it switches its spin orientation by changing the applied magnetic field direction we performed magnetic field cooling experiments under parallel (positive) and antiparallel (negative) magnetic field applied along the c-axis of the chromia film in the region highlighted (dashed white square) in **Fig. 4c**. The results are displayed in **Fig. 5a** and **Fig. 5b** for antiparallel (-0.4 T) and parallel (+ 0.4 T) field to the surface of the chromia film respectively. The lateral width of the edge domains is ~ 330 nm (**Fig. 5c**), consistent with the size of the FIB edge shown in **Fig. 4c**. This makes the width of the edge domain way above the width of DWs (< 100 nm) in **Fig. 3c** and **Fig. 4a**. We explain the magnetic behavior of the edge by the reduction of neighboring spins paired with the creation of defects in the edge region where material has been removed through ion milling. This makes the edge region prone to increased magnetic susceptibility where magnetization is induced with magnetic fields as low as 0.4T while the rest of the long range ordered sample remains in its more robust AFM state which is known to require significantly higher magnetic field to reverse.[43,44] The width of this edge region is therefore not a characteristic length scale associated with domain walls formed by competition between anisotropy and intrinsic exchange but rather a fingerprint of lateral extends of damage created by the ion milling process. In contrast to these edge states, domain walls reflect the intrinsic magnetic properties of a sample in conjunction with extrinsic anisotropy contributions such as strain. Our findings regarding DW width away from artificially created edges is very similar to the DW widths (range of 20–100 nm) reported in references [25] and [35], depending on the type of the DW (Néel or Bloch).[25] This confirms the notion that the DW widths are significantly smaller than expected in the bulk of the low anisotropy magnet $Cr_2O_3$ and determined by epitaxial strain.[46] Recent NV measurements on DWs created via magnetoelectric field cooling in $Cr_2O_3$ single crystals revealed DW-pinning phenomena at mesa edges engineered by electron beam lithography.[35] The observed interaction is explained by the competition between surface energy of the domain wall.

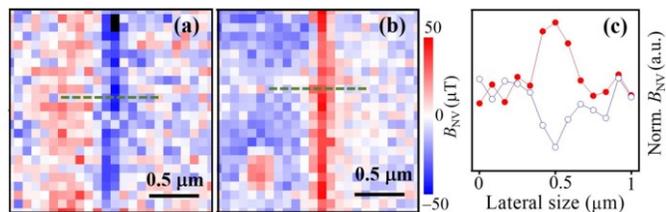

**Figure 5.** $B_{NV}$ map for -0.4 T **(a)** and +0.4 T **(b)** field cooled AFM states showing a magnetic field reversible edge DW. **(c)** Transverse cut of the $B_{NV}$ maps in **(a)** and **(b)** showing the spatial profile of the edge DW.

Of notice is the spatial resolution of our NV-SPM microscope (~ 50 nm), as determined from the magnetization reconstruction of $B_{NV}$ map in **Fig. 3d**, and thus high enough to measure sub-50 nm DWs. We explain the degradation of the spatial resolution from < 15 nm to ~ 50 nm by the large amplitude of the diamond tip oscillation and low Q-factor of the tuning fork cantilever at ambient conditions.[25,21,20] Integrating NV-SPM with ultra-high vacuum system will enhance the Q-factor of the diamond probe by orders of magnitude[20,47] and allow imaging AFM spin textures with a spatial resolution below 15 nm, defined mainly by the distance from the NV tip.[20,21,28,39]

## 4. Conclusions

In Summary, NV stray-field imaging measurements on magnetic field cooled epitaxial $Cr_2O_3$ films revealed switching of the AFM order to just one magnetic state we explained by the presence of a parasitic magnetic moment, tied to Néel vector. The domain formation in PLD grown samples is different from the domain formation in sputtered samples where structural grains act as nucleation centers giving rise to an increased number of domains with reduced size. At the same time, the domain wall width seems to be unaffected by the growth method and is primarily controlled by a narrowing mechanism caused by lattice strain. The lattice strain depends on the mismatch between substrate and thin film and, hence is identical for all $Cr_2O_3$ thin films grown on *c*-plane sapphire. Furthermore by nanostructuring the chromia film we created edge states that switch its orientation based on the direction of applied field during field cooling experiments. These results could motivate further investigations of DW engineering in antiferromagnets.[4,17,25,35] For example boron doping in epitaxial $Cr_2O_3$ films increases $T_N$ to 400 K[48] and enables voltage control of AFM order and Néel vector orientation at zero applied magnetic field.[49] In such structures the AFM spin textures are not yet explored at the nanoscale.

**Author Contributions**
The manuscript was written through contributions of all authors. All authors have given approval to the final version of the manuscript. AE performed NV-SPM measurements and analyzed the data; SQAS and AM grew $Cr_2O_3$ films and performed topography and XRD measurements; RT assisted AE in FIB and SEM measurements; IF wrote the Mathematica code to get the magnetization profile from NV stray field measurements. AL and CB designed the experiments and supervised the project.

**Conflicts of interest**

The authors declare no competing financial interest

**Acknowledgments**

This material is based upon work supported by the NSF/EPSCoR RII Track-1: Emergent Quantum Materials and Technologies (EQUATE), Award OIA-2044049. IF acknowledges support from ERAF project 1.1.1.5/20/A/001. The research was performed in part in the Nebraska Nanoscale Facility: National Nanotechnology Coordinated Infrastructure and the Nebraska Center for Materials and Nanoscience (and/or NERCF), which are supported by NSF under Award ECCS: 2025298, and the Nebraska Research Initiative.

**Corresponding Authors:** *binek@unl.edu, &alaraoui2@unl.edu